\magnification=1200 \vsize=8.9truein \hsize=6.5truein \baselineskip=0.6truecm
\parindent=0truecm \nopagenumbers \font\scap=cmcsc10 \hfuzz=0.8truecm
\parskip=0.2truecm
\font\tenmsb=msbm10
\font\sevenmsb=msbm7
\font\fivemsb=msbm5
\newfam\msbfam
\textfont\msbfam=\tenmsb
\scriptfont\msbfam=\sevenmsb
\scriptscriptfont\msbfam=\fivemsb

\def\yup{y_{n+1}}
\def\xup{x_{n+1}}
\def\x{x_{n}}
\def\y{y_{n}}

\def\xdo{x_{n-1}}

\null \bigskip  \centerline{\bf LINEARISABLE MAPPINGS AND THE LOW-GROWTH
CRITERION}

\vskip 2truecm
\bigskip
\centerline{\scap A. Ramani}
\centerline{\sl CPT, Ecole Polytechnique}
\centerline{\sl CNRS, UMR 7644}
\centerline{\sl 91128 Palaiseau, France}
\bigskip
\centerline{\scap B. Grammaticos}
\centerline{\sl GMPIB, Universit\'e Paris VII}
\centerline{\sl Tour 24-14, 5$^e$\'etage, case 7021}
\centerline{\sl 75251 Paris, France}
\bigskip
\centerline{\scap S. Lafortune$^{\dag}$}
\centerline{\sl LPTMC et GMPIB,  Universit\'e Paris VII}
\centerline{\sl Tour 24-14, 5$^e$\'etage, case 7021}
\centerline{\sl 75251 Paris, France}
\footline{\sl $^{\dag}$ Permanent address: CRM, Universit\'e de Montr\'eal,
Montr\'eal, H3C 3J7 Canada}
\bigskip
\centerline{\scap Y. Ohta}
 \centerline{\sl Department of Applied Mathematics}
\centerline{\sl Faculty of Engineering, Hiroshima University}
 \centerline{\sl1-4-1 Kagamiyama, Higashi-Hiroshima, 739-8527 Japan}
\bigskip\bigskip

Abstract
\smallskip \noindent  We examine a family of discrete second-order systems
which are integrable through reduction to a
linear system. These systems were previously identified using the
singularity confinement criterion. Here we analyse them
using the more stringent criterion of nonexponential growth of the degrees
of the iterates. We show that the linearisable
mappings are characterised by a very special degree growth. The ones
linearisable by reduction to projective systems
exhibit zero growth, i.e. they behave like linear systems, while the
remaining ones (derivatives of Riccati, Gambier
mapping) lead to linear growth. This feature may well serve as a detector
of integrability through linearisation.

\vfill\eject

\footline={\hfill\folio} \pageno=2

Integrability of discrete systems is a concept that can be understood on
the basis of our experience  on
integrable continuous systems. The progress accomplished in the domain of
discrete systems this last decade has made
possible the identification of the possible types of integrability. The
parallel with continuous  systems is almost
perfect. Three main types of integrable discrete systems seem to exist [1]:

a) Systems which possess a sufficient number of constants of motion. The
QRT family of mappings [2] is a nice example of
such a system.

b) Systems which can be reduced to linear mappings. They will be examined
in detail in this paper.

c) Systems which can be obtained as the compatibility condition for some
linear system i.e. systems that possess a Lax
pair. Nice examples of such systems are the discrete Painlev\'e  equations
[3]. Given the Lax pair one can reduce the
integration of the nonlinear mapping to the solution of an  isomonodromy
problem.

It is clear that the integration of a given integrable discrete system may
proceed along any of the lines  sketched
above. One can, for example, perform one first integration using a constant
of motion whereupon the system becomes
linearisable and so on.

The very existence of integrable mappings (and their relative rarity) made
their detection  particularly interesting.
Integrability detectors must, of course, be based on the properties  which
are characteristic of integrability. In this
spirit we have proposed the singularity confinement property [4] based on
the observation that a singularity
spontaneously appearing in an  integrable mapping disappears after some
iterations: it is ``confined'' in the sense that
it does not  propagate {\sl ad infinitum}. The singularity confinement
criterion is a necessary one for integrability
but, as we have already remarked in [1], it is not sufficient. This was
explained in ample details  by Hietarinta and
Viallet [5] who have proposed the notion of algebraic entropy as a stronger
criterion which could well be sufficient.
This criterion is based on the ideas of Arnold [6] and Veselov [7] on the
growth of the degrees of the iterates of some
initial data under the action of the mapping. The main argument is that a
generic, nonintegrable mapping has an
exponential degree growth, while integrability is associated with low
growth, typically polynomial. Although the degree
itself is not invariant under coordinate changes,  the type of growth, as
pointed out by Bellon and Viallet [8], is
invariant. The authors of [5] and [8] have introduced the notion of
algebraic entropy defined as $E=\lim_{n\rightarrow
\infty} {(\log{d_n})/n}$, where $d_n$ is the degree of the $n$th iterate.
Generic, nonintegrable mappings have nonzero
algebraic entropy. The conjecture is that integrability, associated to
polynomial growth, leads to zero algebraic
entropy. In [9] we have examined the results on discrete Painlev\'e
equations based on the singularity confinement
criterion in the light of the low-growth approach. Our main finding was
that singularity confinement is sufficient in
order to deautonomize a given integrable autonomous  mapping. This result
led to the proposal of a dual approach for the
study of discrete integrability based on the successive applications on the
singularity confinement and low-growth
criteria, the latter being implemented only after the first is used to
simplify the problem down to tractable proportions.

The aim of this paper is to examine this particular class of mappings which
are linearisable and study their growth
properties. Most of these systems were obtained using the singularity
confinement criterion and thus a study of the growth
of the degree of the iterates would be an interesting complementary
information. Moreover, as we will show, the
linearisable systems do possess particular growth properties which set them
apart from the other integrable discrete
systems.

The first mapping we are going to treat is a two-point mapping of the form
$x_{n+1}=f(x_n,n)$ where
$f$ is rational in $x_n$ and analytical in $n$. In [1] we have shown that
for all $f$'s of the form
$\sum_i {\alpha_i\over (x+\beta_i)^{\nu_i}}$ the singularity confinement
requirement is satisfied. However all those
mappings cannot be integrable: the discrete Riccati,
$\xup=\alpha+{\lambda\over \x+\beta}$, is the only expected
integrable one. Our argument in [1], for the rejection of these confining
but nonintegrable cases, was based on the
proliferation of the preimages of a given point. If we solve the  mapping
for $x_n$ in terms of $x_{n+1}$ we do not find
a uniquely defined $x_n$ and, iterating, the number of $x_{n-k}$ grows
exponentially. In what follows we shall analyse
this two-point mapping in the  light of the algebraic entropy approach. We
start from the simplest case which we expect
to be nonintegrable,
$$
\xup=\alpha+{\lambda\over \x+\beta}+{\mu\over \x+\gamma}.
\eqno(1)
$$ The initial condition we are going to iterate is
$x_0=p/q$ and the degree we calculate is the homogeneous  degree in $p$ and
$q$ of the numerator (or the denominator) of the iterate. We obtain readily
the following degree sequence
$d_n=1,2,4,8,16,\dots$ i.e. $d_n=2^n$. Thus the algebraic entropy of the
mapping is $\log(2)>0$, an indication that the
mapping cannot be integrable. In the present case it was quite easy to
guess an analytical expression for the degree.
What we do in general in order to obtain a closed-form expression for the
degrees of the iterates, is to compute a
sufficient number of them. Then we establish heuristically an expression of
the degree, compute the next few ones and
check that they agree with the analytical expression prediction. Now we ask
how can one curb the growth and make it
nonexponential. It turns out that the only possibilities are
$\lambda\mu=0$ or
$\beta=\gamma$. In either case mapping (1) becomes a homography. The degree
in this case is simply $d_n=1$ for all $n$.
This is an interesting result, clearly due to the fact that the homographic
mapping is linearisable through a simple
Cole-Hopf transformation.

The second mapping we shall examine is one due to Bellon and collaborators [10]
$$
\matrix{
\displaystyle{\xup={\x+\y-2\x\y^2 \over \y(\x-\y)},} \cr
\displaystyle{\yup={\x+\y-2\x^2\y \over \x(\y-\x)}}. }
\eqno(2)
$$ The degree growth in this case is studied starting from $x_0=r$,
$y_0=p/q$ and again we calculate the homogeneous
degree of the iterate in $p$ and $q$, i.e. we set the degree of $r$ to
zero. (Other choices could have been possible but
the conclusion would not depend on these details.) We obtain the degrees
$d_{x_n}=0,2,2,4,4,6,6,\dots$ and
$d_{y_n}=1,1,3,3,5,5,\dots$ i.e. a linear degree-growth. This is in perfect
agreement with the integrable character of the
mapping. As was shown in [11] it does satisfy the unique preimage
requirement and possesses a constant of motion
$k={1-\x\y\over \y-\x}$, the use of which reduces it to a homographic
mapping for $\x$ or $\y$.

The third mapping we are going to study is the one proposed in [1]
$$
\matrix{
\displaystyle{\xup={\x(\x-\y-a)\over \x^2-\y},} \cr
\displaystyle{\yup={(\x-\y)(\x-\y-a) \over \x^2-\y}} }\eqno(3)
$$ where $a$ was taken constant. We start by assuming that $a$ is an
arbitrary function of $n$ and compute the growth of
the degree. We find $d_{x_n}=0,1,2,3,4,5,6,7,8,\dots$ and
$d_{y_n}=1,2,3,4,5,6,7,8,9,\dots$ i.e. again a linear growth. This is an
indication that (3) is integrable for arbitrary
$a_n$ and indeed it is. Dividing the two equations we obtain
${\yup/\xup}=1-{\y/\x}$ i.e. ${\y/\x}=1/2+k(-1)^n$
whereupon (3) is reduced to a homographic mapping for $x$. Thus in this
case the degree-growth has succesfully predicted
integrability.

A picture starts emerging at this point. While in our study of discrete
Painlev\'e equations and the QRT mapping we found
quadratic growth of the degree of the iterate, linearisable second-order
mappings seem to lead to slower growth. In order
to investigate this property in detail we shall analyse the three-point
mapping we have studied in [12,13] from the point
of view of integrability in general and linearisability in particular. The
generic mapping studied in [13] was one
trilinear in $\x$,
$\xup$, $\xdo$. Several cases were considered. Our starting point is the
mapping,
$$
\xup \x \xdo +\beta \x\xup+\zeta\eta\xup\xdo+\gamma \x\xdo+\beta\gamma
\x+\eta \xdo+\zeta \xup+1=0.
\eqno(4)
$$
We start with the initial conditions $x_0=r$, $x_1=p/q$ and compute the
homogeneous degree in $p$, $q$ at every $n$.
We find
$d_n=0,1,1,2,3,5,8,13,\dots$ i.e. a Fibonacci sequence
$d_{n+1}=d_n+d_{n-1}$ leading to exponential growth of $d_n$ with
asymptotic ratio ${1+\sqrt{5}\over 2}$. Thus mapping (4) is not expected to
be integrable in general. However, as shown
in [13] integrable subcases do exist. We start by requiring that the degree
growth be less rapid and as a drastic decrease
in the degree we demand that $d_3=1$ instead of $2$. We find that this is
possible when either $\beta=\zeta=0$ in which
case the mapping reduces to:
$$
\xup=-\gamma-{\eta \over \x}-{1\over \x\xdo}
\eqno(5)
$$
or $\gamma=\eta=0$, giving a mapping identical to (5) after $x\rightarrow
1/x$. In this case the degree is $d_n=1$ for
$n>0$. Equation (5) is the generic linearisable three-point mapping,
written in canonical form. Its linearisation can be
obtained in terms of a projective system [13] i.e. a system of three linear
equations, a fact which explains the constancy
of the degree.

The trilinear three-point mapping possesses also many nongeneric subcases,
some of which are integrable. The first
nongeneric case writes:
$$
\x(\gamma\xdo+\epsilon)+(\xup+1)(\eta \xdo+1)=0.
\eqno(6)
$$
The degrees of the iterates of mapping (6) form again a Fibonacci sequence
even in the case $\epsilon=0$ or $\eta =0$.
The only case that presents a slightly different behaviour is the case
$\gamma=0$:
$$ (\xup+1)(\eta \xdo+1)+\epsilon \x=0.
\eqno(7)
$$
In the generic case the degree of the iterate behaves like $d_n=$0, 1, 1,
1, 2, 2, 3, 4, 5, 7, 9, 12, 16, 21, 28, 37,
49,$\dots$ satisfying the recursion relation $d_{n+1}=d_{n-1}+ d_{n-2}$
leading to an exponential growth with asymptotic
ratio
$\left({1\over 2}+\sqrt{23\over 108}\right)^{1/3}+\left({1\over
2}-\sqrt{23\over 108}\right)^{1/3}$. Although the mapping
is generically nonintegrable it does possess integrable subcases. Requiring
for example that $d_4=1$ we obtain the
constraint
$\epsilon=\eta=1$ and the mapping becomes periodic with period $5$. If we
require $d_5=1$, we obtain
$\epsilon=-{\eta_{n+1}}(\eta_n-1)$ and
${\eta_{n+1}}\eta_n{\eta_{n-1}}-{\eta_{n+1}}\eta_n+{\eta_{n+1}}-1=0$,
leading again
to a periodic mapping with period $8$. In these cases, the degree of the
iterates exhibits, of course, a periodic
behaviour. A more interesting result is obtained if we require $d_9<7$. We
find that the condition $\eta=1$ and
$\epsilon$ an arbitrary constant leads to a nonexponential  degree growth
$d_n=0,1,1,1,2,2,3,4,5,6,7,9,10,12,14,15,18,20,22,25,27,30,33,36,39,42,46,49,
\dots$. Although the detailed behaviour
of
$d_n$ is pretty complicated one can see that the growth is quadratic: we
have, for example,
$d_{4m+1}=m(m+1)$ for
$m>0$. Thus this mapping is expected to be integrable and indeed, it is a
member of the QRT family. Its constant of
motion is given by
$$ K=\yup+\y-\epsilon\left({\yup\over \y}+{\y\over
\yup}\right)+\epsilon(\epsilon+1)\left({1\over \y}+{1\over
\yup}\right)-{\epsilon^2\over \y\yup}
$$
where $y_k=x_k+1$.
The second nongeneric case is:
$$
\gamma \x \xdo+\delta \xup\xdo+\epsilon \x+\zeta \xup=0.
\eqno(8)
$$ A study of the degree-growth leads always to exponential growth with
asymptotic ratio ${1+\sqrt{5}\over 2}$, except
when
$\gamma=0$ in which case the degrees obey the recurrence
$d_{n+1}=d_{n-1}+d_{n-2}$. No integrable subcases are expected
for mapping (8). The last nongeneric case we shall examine is
$$
\gamma \x\xdo+\xup\xdo+\epsilon \x+\eta\xdo=0.
\eqno(9)
$$  Again the degree sequence is a Fibonacci one except when $\gamma =0$ or
$\eta =0$, in which case we have the
recursion
$d_{n+1}=d_{n-1}+d_{n-2}$, or when $\epsilon_n=\gamma_n \eta_{n-2}$. In the
latter case the degree-growth follows the
pattern $d_n=0,1,1,2,2,3,3,\dots$ i.e. a linear growth. Thus we expect this
case to be integrable. This is precisely what
we found in [13]. Assuming $\eta \neq 0$ we can scale it to $\eta =1$, and thus
$\epsilon=\gamma$. The mapping can then be integrated to the homography
$(\xdo+1)(\x+1)=ka\xdo$ where $k$ is an
integration constant and $a$ is related to $\gamma$ through
$\gamma_n=-{a_{n+1}/a_{n}}$. Thus in this case mapping (9)
is a discrete  derivative of a homographic mapping.

This leads us naturally to the consideration of the generic three-point
mapping that can be considered as the discrete
derivative of a (discrete) Riccati equation. Let us start from the general
homographic mapping which we can write as
$$
A\x\xup+B\x+C\xup+D=0.
\eqno(10)
$$
where $A,B,C,D$ are linear in some constant quantity $\kappa$.
In order to take the discrete derivative we extract the constant $\kappa$
and rewrite (10) as:
$$
\kappa={\alpha\x\xup+\beta\x+\gamma\xup+\delta\over\epsilon\x\xup+\zeta\x+\eta\x
up+\theta}.
\eqno(11)
$$
Using the fact that $\kappa$ is a constant, it is now easy to obtain the
discrete derivative by downshifting (11)
and subtracting it form (11) above.
Instead of examining this most general case we concentrate on the forms
proposed in [14].
They correspond to the reduction of (11) to the two cases:
$$\kappa=\xup+a+{b\over\x}\eqno(12)$$
$$\kappa={\xup(\x+a)\over\x+b}\eqno(13)$$
Next we compute the discrete derivatives of (12) and (13). We find:
$$\xup=\x+a_{n-1}-a_n-{b_n \over \x}+{b_{n-1}\over \xdo}\eqno(14)$$
and
$$\xup=\x{\xdo+a_{n-1}\over\x+a_n}{\x+b_n \over \xdo+b_{n-1}}\eqno(15)$$

The study of the degree of growth of (14) and (15) can be performed in a
straightforward way. For both mappings we find
the sequence
 $d_n=0,1,2,3,4,5,6,\dots$ i.e. a linear growth just as in the cases of
mappings (2), (3) and the integrable subcases of
(9). If we substitute ${b_{n-1}}$ by
${c_{n-1}}$ in the last term of the rhs of (14) or the denominator of (15)
we find $d_n=0,1,2,4,8,16,\dots$ i.e.
$d_n=2^n$ for $n>0$ unless
$c=b$. Investigating all the possible ways to curb the growth we find for
both (14) and (15) that $c=0$ is
also a possibility to bring $d_3$ down to 3. However a detailed analysis
of this case shows that for $c=0$ we have
$d_n=0,1,2,3,5,8,13,21,\dots$ i.e. a Fibonacci sequence with slower, but
still exponential, growth (i.e. ratio
${1+\sqrt{5}\over 2}$ instead of $2$).

One more family of linearisable discrete systems has been studied in detail
in [15] and [16]. They are what we called
the Gambier mappings which constitute the discretisation of the continuous
Gambier equation [17]. The latter is a system
of two Riccati's in cascade. In the discrete case the Gambier system is
written as two homographic mappings which we
write in canonical form as:
$$\yup={a_n\y+b_n\over\y+1}\eqno(16a)$$
$$\xup={\x\y/d_n+c_n^2\over\x+d_n\y}\eqno(16b)$$
Eliminating $y$ we can also write the discrete Gambier system as a single
three-point mapping for $x$.
The study of the degree growth of (16) is straightforward. We start from
$x_0=r$, $y_0=p/q$ and compute the homogeneous
in $p,q$ degree of (16a) and (16b). Since (16a) is a Riccati its degree
does not grow i.e. we have $d_{y_n}=1$. Given the
structure of (16b) we have $d_{x_{n+1}}=d_{x_n}+d_{y_n}$ and thus
$d_{x_n}=n$. What is interesting here is that the
Gambier mapping exhibits a linear degree-growth independently of the
precise values of $a,b,c,d$. The fact that it can be
reduced to Riccati's in cascade is enough to guarantee its integrability.
On the other hand, if we had asked, (as we have
done in [15]) for the possibility to express the solution as an infinite
product of matrices,even across singularities,
this would have led to constraints on the parameters (which were given in
detail in [16]).

In this work we have examined a class of integrable discrete systems
(mainly three-point mappings) from the point of view
of the degree-growth of the iterates of some initial data. Our study was
motivated from the recent works connecting
slow-growth and integrability. Our present analysis confirms our previous
findings based on the singularity confinement
necessary discrete integrability criterion. But what is more important is
that a relation between the details of
integrability and the degree-growth seems to emerge. In this work we have
found two main types of degree-growth: zero and
linear growth. Zero growth is associated to systems which are linearisable
through a reduction to a projective system.
Linear growth is characteristic of systems which can be reduced to linear
ones although at the price of some more
complicated transformations, usually through the existence of some constant
of motion or, as in the case of the Gambier
mapping, through the solutions of linear equations in cascade. On the other
hand, in our study on discrete Painlev\'e
equations and the QRT mapping we found that quadratic growth was the rule.
These results are, of course, characteristic
of three-point (second-order) mappings and we do not expect the details
concerning the precise exponents to carry over
to higher-order mappings. Still, we expect the pattern detected here,
namely that linearisable mappings lead to
slower growth than the nonlinearisable integrable ones, to persist.
It could be used for the
classification of integrable discrete systems and be a valuable indication
as to the precise method of their
integration. We intend to return to this point in some future work.
\vfill\eject
\noindent {\scap Acknowledgements}.
\smallskip
\noindent  The authors are grateful to J. Fitch who provided them with a
new (beta) version of REDUCE without which the
calculations presented here would have been impossible.
 S. Lafortune acknowledges two scholarships: one from FCAR du Qu\'ebec for
his Ph.D. and one from ``Programme de Soutien
de Cotutelle de Th\`ese de doctorat du Gouvernement du Qu\'ebec'' for his
stay in Paris.
\bigskip
{\scap References}
\smallskip
\item{[1]} B. Grammaticos, A. Ramani, K. M. Tamizhmani, Jour. Phys. A 27
(1994) 559.
\item{[2]} G.R.W. Quispel, J.A.G. Roberts and C.J. Thompson, Physica D34
(1989) 183.
\item{[3]} B. Grammaticos, F. Nijhoff and A. Ramani, {\sl Discrete
Painlev\'e equations}, course at the Carg\`ese 96
summer school on Painlev\'e equations.
\item{[4]}	B. Grammaticos, A. Ramani and V.G. Papageorgiou, Phys. Rev.
Lett. 67 (1991) 1825.
\item{[5]} J. Hietarinta and C.-M. Viallet, Phys. Rev. Lett. 81 (1998) 325.
\item{[6]} V.I. Arnold, Bol. Soc. Bras. Mat. 21 (1990) 1.
\item{[7]} A.P. Veselov, Comm. Math. Phys. 145 (1992) 181.
\item{[8]} M.P. Bellon and C.-M. Viallet, {\sl Algebraic Entropy}, Comm.
Math. Phys. to appear.
\item{[9]} Y. Ohta, K.M. Tamizhmani, B. Grammaticos and A. Ramani, {\sl
Singularity confinement and algebraic
entropy: the case of the discrete Painlev\'e equations}, preprint (1999).
\item{[10]} M.P. Bellon, J.-M. Maillard and C.-M. Viallet, Phys. Rev. Lett.
67 (1991) 1373.
\item{[11]} B. Grammaticos, A. Ramani, Int. J. of Mod. Phys. B 7 (1993) 3551.
\item{[12]} A. Ramani, B. Grammaticos, G. Karra, Physica A 180 (1992) 115.
\item{[13]} A. Ramani, B. Grammaticos, K.M. Tamizhmani, S. Lafortune,
Physica A 252 (1998) 138.
\item{[14]} B. Grammaticos, A. Ramani, Meth. and Appl. of An. 4 (1997) 196.
\item{[14]} B. Grammaticos and A. Ramani, Physica A 223 (1995) 125.
\item{[15]} B. Grammaticos, A. Ramani, S. Lafortune Physica A 253 (1998) 260.
\item{[16]} E.L. Ince, {\sl Ordinary differential equations}, Dover, New
York, 1956.
\end